\magnification 1200
\input amstex

\catcode`\X=11 \catcode`\@=\active
\documentstyle{amsppt}
\refstyle{A}

\NoRunningHeads

\catcode`\X=12\catcode`\@=11
\def\n@wcount{\alloc@0\count\countdef\insc@unt}
\def\n@wwrite{\alloc@7\write\chardef\sixt@@n}
\def\n@wread{\alloc@6\read\chardef\sixt@@n}
\def\crossrefs#1{\ifx\alltgs#1\let\tr@ce=\alltgs\else\def\tr@ce{#1,}\fi
   \n@wwrite\cit@tionsout\openout\cit@tionsout=\jobname.cit
   \write\cit@tionsout{\tr@ce}\expandafter\setfl@gs\tr@ce,}
\def\setfl@gs#1,{\def\@{#1}\ifx\@\empty\let\next=\relax
   \else\let\next=\setfl@gs\expandafter\xdef
   \csname#1tr@cetrue\endcsname{}\fi\next}
\newcount\sectno\sectno=0\newcount\subsectno\subsectno=0\def\r@s@t{\relax}
\def\resetall{\global\advance\sectno by 1\subsectno=0
   \gdef\firstpart{\number\sectno}\r@s@t}
\def\resetsub{\global\advance\subsectno by 1
   \gdef\firstpart{\number\sectno.\number\subsectno}\r@s@t}
\def\v@idline{\par}\def\firstpart{\number\sectno}
\def\l@c@l#1X{\firstpart.#1}\def\gl@b@l#1X{#1}\def\t@d@l#1X{{}}
\def\m@ketag#1#2{\expandafter\n@wcount\csname#2tagno\endcsname
     \csname#2tagno\endcsname=0\let\tail=\alltgs\xdef\alltgs{\tail#2,}%
   \ifx#1\l@c@l\let\tail=\r@s@t\xdef\r@s@t{\csname#2tagno\endcsname=0\tail}\fi
   \expandafter\gdef\csname#2cite\endcsname##1{\expandafter
     \ifx\csname#2tag##1\endcsname\relax?\else{\rm\csname#2tag##1\endcsname}\fi
     \expandafter\ifx\csname#2tr@cetrue\endcsname\relax\else
     \write\cit@tionsout{#2tag ##1 cited on page \folio.}\fi}%
   \expandafter\gdef\csname#2page\endcsname##1{\expandafter
     \ifx\csname#2page##1\endcsname\relax?\else\csname#2page##1\endcsname\fi
     \expandafter\ifx\csname#2tr@cetrue\endcsname\relax\else
     \write\cit@tionsout{#2tag ##1 cited on page \folio.}\fi}%
   \expandafter\gdef\csname#2tag\endcsname##1{\global\advance
     \csname#2tagno\endcsname by 1%
   \expandafter\ifx\csname#2check##1\endcsname\relax\else%
\fi
   \expandafter\xdef\csname#2check##1\endcsname{}%
   \expandafter\xdef\csname#2tag##1\endcsname
     {#1\number\csname#2tagno\endcsnameX}%
   \write\t@gsout{#2tag ##1 assigned number \csname#2tag##1\endcsname\space
      on page \number\count0.}%
   \csname#2tag##1\endcsname}}%
\def\m@kecs #1tag #2 assigned number #3 on page #4.%
    {\expandafter\gdef\csname#1tag#2\endcsname{#3}
    \expandafter\gdef\csname#1page#2\endcsname{#4}}
\def\re@der{\ifeof\t@gsin\let\next=\relax\else
    \read\t@gsin to\t@gline\ifx\t@gline\v@idline\else
\expandafter\m@kecs \t@gline\fi\let \next=\re@der\fi\next}
\def\t@gs#1{\def\alltgs{}\m@ketag#1e\m@ketag#1s\m@ketag\t@d@l p
    \m@ketag\gl@b@l r \n@wread\t@gsin\openin\t@gsin=\jobname.tgs \re@der
    \closein\t@gsin\n@wwrite\t@gsout\openout\t@gsout=\jobname.tgs }
\outer\def\localtags{\t@gs\l@c@l}
\outer\def\globaltags{\t@gs\gl@b@l}
\outer\def\newlocaltag#1{\m@ketag\l@c@l{#1}}
\outer\def\newglobaltag#1{\m@ketag\gl@b@l{#1}}

\def\t@gsoff#1,{\def\@{#1}\ifx\@\empty\let\next=\relax\else
\let\next=\t@gsoff
   \expandafter\gdef\csname#1cite\endcsname{\relax}
   \expandafter\gdef\csname#1page\endcsname##1{?}
   \expandafter\gdef\csname#1tag\endcsname{\relax}\fi\next}
\def\verbatimtags{\let\ift@gs=\iffalse\ifx\alltgs\relax\else
   \expandafter\t@gsoff\alltgs,\fi}
\def\(#1){\edef\dot@g{\ift@gs\if@lign(\noexpand\etag{#1})
    \else\ifmmode\noexpand\tag\noexpand\etag{#1}%
    \else{\rm(\noexpand\ecite{#1})}\fi\fi\else\if@lign{\rm(#1)}
    \else\ifmmode\noexpand\tag#1\else{\rm(#1)}\fi\fi\fi}\dot@g}
\let\ift@gs=\iftrue\let\if@lign=\iffalse
\let\n@weqalignno=\eqalignno
\def\eqalignno#1{\let\if@lign=\iftrue\n@weqalignno{#1}}

\catcode`\X=11 \catcode`\@=\active

\localtags
\TagsOnRight


\define\R{\Bbb R}
\define\real{\Bbb R}

\define\eps{\epsilon}

\define\Z{\Bbb Z}
\define\Prob{\bold P}
\define\E{\bold E}
\define\esssup{\operatornamewithlimits{ess\,sup}}

\title
Phase Segregation Dynamics \\
in Particle Systems
 with Long Range Interactions I:\\
 Macroscopic Limits
\endtitle

\leftheadtext\nofrills
{G.Giacomin and J.L.Lebowitz}
\rightheadtext\nofrills
{A phase segregation model}

\author
\centerline{Giambattista Giacomin}
\centerline{Institut f\"ur  Angewandte Mathematik}
\centerline{Universit\"at Z\"urich--Irchel}
\centerline{Winterthurer Str. 190, CH--8057 Z\"urich Switzerland}
\bigskip
\centerline{Joel L. Lebowitz}
\centerline{Department of Mathematics}
\centerline{Hill Center, Rutgers University}
\centerline{New Brunswick, N.J. 08903, U.S.A.}
\endauthor

\abstract
We present and discuss
the derivation of  a nonlinear non-local integro-differential equation
for the macroscopic time evolution of the conserved order parameter
$\rho({ r}, t)$ of a   binary alloy undergoing phase
segregation.  Our model is a $d$-dimensional lattice gas evolving via
Kawasaki exchange dynamics, i.e.\
a (Poisson)
nearest--neighbor exchange
process, reversible with respect to the Gibbs measure for
a Hamiltonian which includes both short range (local) and long range 
(nonlocal) interactions.
The nonlocal part is given by a  pair  potential
$\gamma^d J (\gamma \vert
x-y \vert)$, $\gamma >0$, $x$ and $y$ in $\Z ^d$, in the limit
$\gamma \to 0$ .
 The
macroscopic evolution is observed on the spatial scale
$\gamma^{-1}$ and time scale $\gamma^{-2}$,
i.e., the density, $\rho({ r}, t)$, is the empirical
average of the occupation numbers over a small
macroscopic volume element
centered at ${ r} = \gamma x$.
A rigorous derivation is presented in the case in which
there is no local interaction.
In a subsequent paper (part II), we discuss the phase segregation
phenomena in the model. In particular
we argue that the phase boundary evolutions, arising
as sharp interface limits of the family of equations
derived in this paper,
are the same as the
ones obtained from the corresponding limits for the
Cahn-Hilliard equation.
\hfill\break
\phantom{a}\hfill\break
\phantom{a}\hfill\break
\endabstract
\endtopmatter

\document

\head
1. Introduction
\endhead
\resetall
\sectno=1

The process of phase segregation, following a quench (sudden cooling) of a
system from  high temperature, where the system has a unique uniform
equilibrium phase, into the miscibility gap where two (or more) phases can
coexist, is variously known as spinodal decomposition, nucleation,
coarsening, etc$\ldots$
It concerns the tendency of the different phases to
segregate, creating  larger and larger
 domains of approximately homogeneous single
phase regions.  The problem is of great practical importance in the
manufacturing of alloys where the degree of segregation influences
the properties of the material.
The mathematical description of the time evolution of
the local macroscopic order parameter in such systems, e.g.\ the
difference in the concentration of A- and B-atoms
in a binary A-B alloy, is
commonly  given by nonlinear fourth order
equations of the Cahn-Hilliard type [\rcite{CH}].

These equations appear to capture much of the phenomena.  In particular,
their numerical solutions show good agreement with experiments
and with computer simulations of the Ising model (thought of as a
binary alloy) evolving via Kawasaki exchange dynamics.  This agreement
relates both to the appearance and shape of segregated
domains, which seem to exhibit a self-similar structure, and also to the
quantitative behavior of the characteristic length describing these
structures,  which seems to
grow like $t^{1/3}$, where $t$ is the time.
While this agreement is certainly satisfying,
 the original Cahn--Hilliard equation
(CHE), or the modifications of it
proposed so far [\rcite{Fife}],  do not seem to
to arise as  exact macroscopic
description of microscopic models of interacting particles,
such as the Ising model with Kawasaki dynamics (the CHE
can however be derived from certain mesoscopic Ginzburg--Landau
continuous spin models [\rcite{BLO}]).
This is unlike some other physically motivated equations, e.g.
 the
{\sl diffusion} equation and the
 {\sl Boltzmann} equation, which can be derived from idealized 
  microscopic
models in suitable limits [\rcite{DP},\rcite{Spohn}].
Such derivations are both of intrinsic
interest
and also indicate something
about the range of
applicability of the macroscopic equations. 
The latter might be particularly relevant for the CHE, where
all that is known mathematically about the behavior of the
solutions is restricted to the late stages of the coarsening process when
the evolution is {\sl assumed} to be dominated by the
motion of sharp interfaces
between well formed domains of the pure phases.
It is far from clear
how much this singles out the CHE from  other possible
equations
describing phase segregation.

In this note we rigorously derive a macroscopic equation, describing phase
segregation in microscopic model systems with long range interactions
evolving according to stochastic
Kawasaki dynamics with nearest neighbor exchanges.  We will then, in
Part II [\rcite{GL2}],
study the interface motion obtained from
the derived macroscopic equation,
in the {\sl sharp interface limit},
by means of formal matched asymptotic
expansions of the solution of the macroscopic evolution
equation (see e.g.
[\rcite{Caginalp},\rcite{CENK},\rcite{Langer},\rcite{Pego}]).
By {\sl sharp interface} we mean the limit in which the
phase
domains are very large compared to the size of the interfacial region, i.e.
denoting by $L$ the {\sl typical }
size of the domains, we  look for results in the limit $L \rightarrow
\infty$. The time will have to be properly scaled as well, typically
as some integer power of $L$, according to the type
of initial condition and the choice of the temperature.  Our conclusion
there is that, from the {\it sharp interface viewpoint}, the equation
derived from a particle system and the Cahn-Hilliard equation are
essentially equivalent.

The models we consider are dynamical versions
of lattice gases interacting via long range Kac potentials,
also known as {\sl
local mean field} interactions.
The equilibrium properties of  these systems are well known
([\rcite{Kac},\rcite{LP},\rcite{PL}]).
They provide microscopic models in which the van der Waals or mean field
description of
phase transition phenomena, which is
in good qualitative (or even quantitative) agreement with experiments away
from the critical point, holds exactly.  This includes
 metastability phenomena [\rcite{PL}].
The corresponding dynamical
 models which we study here are sometimes called {\sl local mean field
Kawasaki dynamics}.
They can be described in words as follows:
each particle hops (at a random  time)
 from a site of the lattice ${\Bbb Z} ^d$ to one of its $2d$ neighboring
sites with a rate
which depends on the
particle configuration. These rates are chosen to satisfy detailed balance
(reversibility) with respect to the Gibbs measure having the specified
interaction
 between the particles [\rcite{Spohn}].

In the simplest model we consider here, there is only
a long range (Kac type)
potential.
We will also discuss, but not investigate in detail, the case in which
 additional short range
interactions are also present.  Further work on the same model and
related ones can be found in [\rcite{G1},\rcite{LOP},\rcite{Yau}]:
these papers focus on the diffusive regime, i.e.
on the region in which there is only one phase, but versions
of the integral equation on which we are focusing are already
present there.
The case without a conservation law, {\sl Glauber} dynamics,
has received much attention, see [\rcite{DOPT}] and references
therein.

The precise definition of the model is given
in the next section.
We show there that the evolution of the
macroscopic density (in the simplest model) is given in terms of
 a second order integro-differential equation \(maineq).
We then show that this equation can be written in terms of the gradient flux
associated with the classical local mean field free energy functional
([\rcite{LP},\rcite{PL}])
 and a density dependent  mobility.
This  allows us to make a direct connection between
the properties of the solution of the derived evolution equation and the
equilibrium phase diagram as well as with the solution of the CHE.
In section 3
we argue that the gradient structure for the macroscopic
evolution law should hold generally in systems with long range Kac
potentials and arbitrary additional short range interactions.
In Section 4 we make a remark on the
case in which a weak external field is present.
 The proof for the
simple case is given in section 5.

\head
2. The particle model and its hydrodynamic limit
\endhead
\resetall
\sectno=2

The particles  live on the $d$--dimensional lattice $\Lambda _\gamma=
\{1, 2, \ldots ,[\gamma ^{-1}] \}^d$, where $\gamma>0$ is a small
parameter and $[r]$ denotes the integer part of the real number $r$.
We impose periodic boundary conditions on $\Lambda _\gamma$.
Each site of $\Lambda _\gamma$ is either occupied (1) or empty (0),
hence
a particle configuration is an element $\eta $ of
$\Omega _\gamma =\{0,1\}^{\Lambda _\gamma}$ and the latter
is endowed with the product topology.
The dynamics is specified by giving an initial condition
$\eta _0 \in \Omega _\gamma$, which may be random, i.e. a measure
on the Borel sets of $\Omega _\gamma$,
and some (stochastic) evolution rules which will
define $\eta _t$ for any $t$ positive. Our aim is to
have a Markovian dynamics for which the Gibbs measure associated
with a given  Hamiltonian and a given total particle number is the unique
reversible time invariant measure at a fixed
temperature.
The Hamiltonian is a real valued function defined on $\Omega_\gamma$
and we take it to be the sum of two terms:
$$
H  =
H_{s} +H_{\gamma}
\(Hamilt)
$$
$$
H_s=
-{1\over2} \sum _{x, y \in \Lambda _\gamma}
K(x -y) \eta (x) \eta(y)
\(HL)
$$
$$
H_\gamma =
 -{1\over 2}
\sum_{x, y \in \Lambda _\gamma}
\gamma ^d J(\gamma (x-y)) \eta (x) \eta (y)
\(HNL)
$$
in which $J$
is a smooth function ($C^\infty$) from the
$d$--dimensional unit torus $T^d$ to the real numbers,
such that $J(r)=J(-r)$,  and
$K( x )=0$ if $\vert x\vert >R$, for some
$R$ independent of $\gamma$.
 The term in \(HNL) will
be called {\sl nonlocal}, while the one in \(HL) will be
called {\sl local} or {\sl short range}.
The Gibbs measure with Hamiltonian
$H_\gamma$ at the temperature $1/\beta$ ($\beta>0$)
and total number of particle $N\in {\Bbb Z} ^+$  is defined
as
$$
\mu^\beta_\gamma(\eta )={ \exp \left(-\beta H (\eta) \right)
\over Z_\gamma(N)}
\(Gibbs)
$$
where
$$
Z_\gamma (N) =
\sum _{\eta \in {\Omega _\gamma^N}}
\exp \left(-\beta H(\eta ) \right).
\(Zgamma)
$$
\(Gibbs)  is a probability measure over $\Omega_\gamma^N
\equiv
\{\eta \in \Omega_\gamma :
\sum_{x \in \Lambda_\gamma} \eta(x)=N\}$.
The stochastic process $\{ \eta _t\}_{t\ge 0}$
is the Poisson jump process [\rcite{Liggett},\rcite{Spohn}]
generated by the operator
$$
L_\gamma f(\eta)=
\sum_{x,y \in \Lambda_\gamma}
c_\gamma (x,y; \eta)
\left[f(\eta ^{x,y}) -f(\eta)\right]
\(generator)
$$
where $f$ is a real valued (bounded) function on $\Omega_\gamma$,
$$
\eta ^{x,y} (z)=
\cases
\eta(x) & \text{if } z=y \\
\eta(y) & \text{if } z=x \\
\eta(z) &\text{otherwise}
\endcases
\(exchangexy)
$$
and
$$
c_\gamma (x,y; \eta)=
\Phi \left\{
{\beta }
\left[H (\eta^{x,y})-H (\eta )\right]
\right\}
\(poissonrate)
$$
if $\vert x-y\vert =1$ and is zero otherwise.
Here $\Phi: \real \rightarrow \real ^+$ is twice differentiable in
a neighborhood of $0$
and satisfies the detailed balance or reversibility condition,
$$
\Phi(E)= \exp(-E)\Phi(-E)
\(detbalance)
$$
 for all $E\in \real$. We also assume $\Phi(0)=1$: the case
$\Phi (0)\in (0, \infty)$ can be recovered with a time change.

Loosely speaking, the process $\eta_t$ can be described in
the following way:
if at time $t$ the configuration is $\eta _t$, the probability
that in the time interval $[t,t+\Delta t]$ the sites $x,y$
($\vert x-y \vert =1$) exchange their occupation numbers  is
$$
c_\gamma (x,y; \eta _t) \Delta t + O\left(
(\Delta t)^2 \right).
\(deltaprob)
$$
We note that if $\eta(x)=\eta(y)$, then an exchange
between $x$ and $y$
does not modify the configuration $\eta$ and it is thus
possible to interpret the dynamics in terms of
particles which attempt to jump from $x$ to $y$, but
the jump is performed only if the site $y$ is empty: this
type of dynamics is said to have an {\sl exclusion} rule,
that is the particles have an on site hard--core repulsion.
A detailed construction of this process
in terms of Poisson jump processes is given
on page 158 of [\rcite{Spohn}]. The configuration
space we are working on ($\Omega_\gamma$) is finite and
this avoids the difficulties connected to
defining such a dynamics on an infinite state space
(Chapter 1 of [\rcite{Liggett}]). In Section 1.2 (part II)
of [\rcite{Spohn}] it is shown
that for $f$ and $g$, bounded functions on $\Omega_\gamma$
$$
\int
g(\eta) L_\gamma f(\eta) \text{  d} \mu^\beta_\gamma (\eta)=
\int
f(\eta) L_\gamma g(\eta) \text{ d} \mu^\beta _\gamma (\eta)
\(reversible)
$$
This property is called {\sl reversibility}
and it is a direct consequence of
\(detbalance).
In particular \(reversible) implies
that
$$
\int
L_\gamma f(\eta) \text{ d} \mu^\beta_\gamma (\eta)=
0
\(invariant)
$$

We will use the following notation:
the generic initial condition is a probability measure $\mu$
on $\Omega_\gamma$.
The law of the process $\{ \eta _t\}_{t \ge 0}$
with initial condition $\mu$
will be denoted by $P_\gamma ^{\mu}$ ($E_\gamma ^{\mu}$ for
the expectation). The process $P_\gamma ^\mu$ is linked
to the semigroup generated by $L_\gamma$ via the formula
$E_\gamma ^{\mu} (f(\eta_t))=
\int (\exp(L_\gamma t) f )(\eta) \mu (\text{d} \eta)$.
Equation \(invariant) implies that for any $t \ge 0$ and
any $f$ bounded
$$
{\text{ d}\over \text{ d}t}
E_\gamma ^{\mu^\beta_\gamma} \left(
f(\eta_t) \right)=
0
\(inv2)
$$
which means that $\mu_\gamma ^\beta$ is invariant
 under the process generated by $L_\gamma$.

\definition{The hydrodynamic limit}
We are interested in initial states $\mu_\gamma$ such that,
when $\gamma \rightarrow 0$, $\mu_\gamma$ {\sl resembles}
more and more a profile $\rho_0 $, where $\rho_0$
is a measurable function from the $d$-dimensional unit terms
$T^d$ to $[0,1]$, stretched by $\gamma ^{-1}$.
More precisely, we say that $\{\mu_\gamma\}_{\gamma >0}$
is an initial condition associated with $\rho_0$ if, for
any continuous function $\phi$ from $T^d$ to $\real$ and every $\delta >0$,
$$
\lim_{\gamma \rightarrow 0}
\mu_\gamma \left(
\left \vert
\gamma ^d \sum_{x \in T^d} \phi (\gamma x) \eta (x)-
\int_{T^d} \phi (r) \rho_0(r) \text{d}r
\right\vert >\delta
\right) =0.
\(initprob)
$$

The condition \(initprob) is clearly satisfied if $\mu_\gamma$
is such that $\langle \eta (x) \rangle =
\int _{\Omega _\gamma} \eta (x) \text{d} \mu_\gamma (\eta)$ $=
\rho_0(\gamma x)$ for all $x $ in $ \Lambda _\gamma$ and
the occupation number of the sites are independent.
On the other hand, the initial condition concentrated
on the chessboard configuration ($\eta (x)=1$ if the sum of the components
of $x$ is even, $\eta(x)=0$ otherwise) is also obviously associated
with $\rho_0\equiv 1/2$ and many other examples can be easily
constructed. Loosely speaking, when we say that the particle system
has a hydrodynamic limit, we mean that
 \(initprob)
holds also at later times, if we replace $\rho_0$ by
the solution   of a suitable
hydrodynamic evolution equation (in our
case we will have an integro-differential equation),
with initial condition $\rho_0$. In Section 4 we prove
the following Theorem:
\enddefinition

\proclaim{Theorem 1}
{\sl The Hydrodynamic limit without short range interactions}.
Set $K(x) $ $ =0$
so that  the Hamiltonian
coincides with $H_\gamma$.
Let $\mu_\gamma$ be an initial
condition associated with $\rho_0\in C^2$.
Then, for any $t$ positive , $\delta >0$, and any continuous
function $\phi$ from $T^d$ to $\real$
$$
\lim_{\gamma \rightarrow 0}
P_\gamma^{\mu_\gamma}
\left(
\left \vert
\gamma ^d \sum_{x \in T^d} \phi (\gamma x) \eta _{t \gamma ^{-2}} (x)-
\int_{T^d} \phi (r) \rho(r,t) \text{d}r
\right\vert >\delta
\right) =0
\(hydrolim)
$$
when $\rho(r,t)$, $r \in T^d$ and  $t \in [0, \infty)$,
 is the unique classical solution
 of the equation
$$
\cases
\partial _t\rho(r,t)=
\nabla\cdot \left[
\nabla \rho(r,t)-\beta \rho(r,t)(1-\rho(r,t))
\int _{T^d} \nabla J (r-r^\prime ) \rho (r^\prime) \text{d}r^\prime
\right] & \\
 & \\
\rho(r,0)=\rho_0(r)&
\endcases
\(maineq)
$$
\endproclaim

\vskip 0.3 cm

It is now an observation, which at first sight appears surprising, that
\(maineq) can be rewritten in the form
$$
{\partial \rho({ r},t) \over \partial t} = \nabla \cdot
\left[
\sigma^0(\rho)\left(\nabla {\delta \over \delta \rho} {\Cal F}^0\right)
\right]
\(newone)
$$
where
$$
{\Cal F}^0(\rho)=
-{1\over \beta}\int_{T^d} s(\rho(r)) \text{d}r -{1\over 2}
\int \int _{T^d\times T^d}
J( r- r^\prime  ) \rho (r) \rho (r^\prime)
\text{d} r \text{d} r^\prime
\(freeE)
$$
with
$$
s(\rho)=-\rho \log \rho -(1-\rho) \log(1-\rho).
\(entropy)
$$
and 
$$
\sigma ^0(\rho) \equiv \beta\rho (1-\rho)
\(sigma)
$$

Rewriting \(maineq) in the form \(newone),  ${\Cal F}^0$ is
recognized as the free energy functional and $\sigma^0 (\rho)$ as the
mobility  of our model without the long range interactions.  This  allows us to connect
our equation with that of the CHE.  Before doing that we
rewrite ${\Cal F}^0(\rho)$, up to an irrelevant additive constant,
 in the form
$$
{\Cal F}^0(\rho) = \int _{T^d} f_c^0(\rho(r)) \text{d} r
+
{1 \over 4}
\int_{T^d \times T^d} J(r-r^\prime)
\left[\rho (r)- \rho (r^\prime)\right]^2 \text{d}r \text{d}r^\prime
\(newway)
$$
where
$$
f_c^0 (\rho)=- {{\hat J}(0)\over 2} \left(
\rho -{1\over 2} \right)^2
+
{1\over \beta }
\left(
\rho \log \rho +(1-\rho) \log (1-\rho)
\right)
\(fbeta)
$$
and ${\hat J}(0)=\int J(r)\text{d}r$.

Note that if $\rho$ is constant the second term in
\(newway) vanishes, so that $f_c^0$
is the {\sl constrained equilibrium} free energy density of a homogeneous
system see
[\rcite{LP}, \rcite{PL}]).
$f_c^0(\rho)$ is in fact the correct equilibrium free energy density as long
as $\beta \leq \beta_c = 1/T_c = 4/{\hat J}(0)$.  If $\beta > \beta_c$,
 $f^0_c(\rho)$ has a double minimum at $\rho = \rho^{\pm}_\beta$, the two
nontrivial solutions of $\log (\rho/(1 - \rho)) = \beta {\hat J}(0)(\rho -
1/2)$, and the correct free energy is then obtained by the double tangent
construction.
 A heuristic derivation  of \(maineq) and an
explanation of \(newone)  is given
in Section 3, where the generalization to the case $K \not=0$ is
considered.  \(newone) has
the same structure as the CHE, whose various forms correpond to different
$\sigma^0 (\rho)$ and $f_c^0 (\rho)$ chosen  by
different authors [\rcite{CH},\rcite{CENK},\rcite{Langer},\rcite{Puri}].
  What is common to the different CHE's in the
literature is that the second term on the right side of \(newway) is of the
form ${1 \over 2} \zeta \int (\nabla \rho)^2 \text{d}
{ r}$ where $\zeta>0$ is
related to the surface tension.  This can be thought of as expanding the
term in \(newway) and keeping only some terms, which is reasonable when
the scale on which $\rho$ varies is large compared to $\gamma^{-1}$.  We shall discuss the
relationship between the solutions of \(maineq) and the CHE in
[\rcite{GL2}], see also [\rcite{GLprl}].

\vskip 0.3 cm
\noindent

\head
3.  The general case: $K \not=0$
\endhead
\resetall
\sectno=3

We start by giving a heuristic explanation
of the result in Theorem 1, which is in fact a sketch
of its proof. We are for the moment still in the case $K=0$:
if $\beta=0$ the particle system reduces to
the symmetric {\sl simple exclusion process} (SEP),
i.e. all the particles are performing exchanges with rate
one, so their only interaction is given
by the exclusion rule.
As it is straightforward to verify, the Bernoulli measures $\mu_\rho$
with uniform
density $\rho \in [0,1]$, under which
the random variables $\{\eta(x)\}_{x \in \Lambda_\gamma}$ are
independent and $\int \eta(0) \text{d}\mu_\rho(\eta)=\rho$,  are invariant
for the SEP dynamics and the hydrodynamic limit
for the SEP is simply given
by the heat equation (see e.g. [\rcite{KOV}]),
as in \(maineq) with $\beta=0$.
Moreover, we observe that in the case $\beta \ge 0$,
$$
\align
H (\eta ^{x, x+e}) -H(\eta) &=
{\gamma }
\left( \eta (x+e) -\eta(x)\right)
\left[
\gamma ^d
\sum_z \eta (z) (e\cdot \nabla J) (\gamma (x-z)) \right]+O(\gamma ^2)
\\
& = O(\gamma)
\(firstobserve)
\endalign
$$
for all $x \in \Lambda _\gamma$ and $e$ a unit vector
in ${\Bbb Z}^d$ (\(firstobserve) is
derived in  Lemma 2, Section 5), so that
$$
\align
c_\gamma
 (x, x+e; \eta)&=1-{\beta \over 2}\left[H (\eta ^{x, x+e}) -H(\eta)
\right] +O(\gamma ^2)
\\
& = 1+O(\gamma)
\(expandc)
\endalign
$$
where we used \(firstobserve) and
the fact that \(detbalance) and $\Phi (0)=1$
imply $\Phi ^\prime (0)=-1/2$ .

Formula \(expandc) indicates clearly that the dynamics with $\beta>0$
is a weak perturbation of the $\beta =0$ dynamics.
In particular the dominant SEP dynamics will enforce at time
$t \gamma ^{-2}$ ($t>0$)  {\sl local equilibrium} with respect
to its invariant measure, i.e. the state of the
system will locally (on spatial scales
shorter that $\gamma ^{-1}$) be very close
to the Bernoulli measure $\mu_{\rho}$ and $\rho $ will be varying on the
macroscopic
scale $\gamma ^{-1}$.
The perturbation term in \(firstobserve) generates a force term given by
the negative gradient of the energy density at $r$,
$$
F(r)\equiv \nabla
\left(
\int J(r-r^\prime) \rho(r^\prime) \text{d}r^\prime\right)
\(force)
$$
This gives an extra contribution to the macroscopic current
equal to this force times the mobility $\sigma^0=\beta \rho (1-\rho)$.
In this expression $\beta$ measures the intensity of the bias
in the exchange rates and $\rho(1-\rho)$ gives the rate
at which the exchanges
actually take place when a particular site is chosen at random,
since the system is locally described by the Bernoulli measure
$\mu_\rho$. This explains the form of \(maineq).

That \(maineq) can be transformed into \(newone)
is clearly due to the fact that $s(\rho)$ in \(entropy) satisfies the
relation,
$$
- {1\over \beta}
 s^{\prime\prime}(\rho) \sigma ^0 (\rho)
=1
\(tobeclar)
$$
for all $\rho \in (0,1)$. This is no accident but, as will now show,
\(tobeclar)
follows from  the {\sl Einstein relation} between fluxes and forces
for this system.

This will become clearer if we consider the general
case $K\not=0$.  We call the system with only the
local Hamiltonian $H_s$ the {\sl reference} system.
The equilibrium
free energy
$f^s_{eq}(\rho)$
associated with this reference system at an average density $\rho
\in [0,1]$, which depends also on $\beta$, is uniquely defined
in {\sl the thermodynamic limit}
(see [\rcite{Ruelle}, \rcite{Simon}]).
We shall further assume that
$f^s_{eq }(\rho)$ is  strictly convex and real analytic,
which implies that there is no phase transition
for the equilibrium reference system, associated with $H_s$ at the temperature
$1/\beta$. Morever we are assuming that for each $\rho\in
[0,1]$ there exists
a unique, translation invariant
and ergodic,  infinite volume limit Gibbs state
$\mu^s_\rho$ such that $\int \eta (0) \text{d}
\mu^s _\rho (\eta)=\rho$.
All these properties, which are to a certain extent equivalent,
are known to hold
if $\beta $ is sufficiently small (see
[\rcite{Ruelle}, \rcite{Simon}]).

We now claim that the correct macroscopic evolution law
for the case $K\not=0$ should be 
$$
{\partial \rho \over \partial t} =
\nabla \cdot\left[
\sigma_s \nabla
\left(
{\delta {\Cal F} \over \delta \rho }
\right)
\right]
\(generaleq)
$$
where  $\sigma_s(\rho)$ is the mobility of the reference system,
$$
{\Cal F} (\rho) =
\int f_c (\rho (r)) \text{d} r+
{1\over 4} \int \int J(r-r^\prime) \left[
\rho (r) -\rho(r^\prime) \right]^2
\text{d} r \text {d} r^\prime
\(newway)
$$
and
we have defined
the constrained equilibrium free energy density (in analogy with [\rcite{LP}])
as
$$
f_c (\rho) = - {{\hat J}(0) \over 2}\left(
\rho -{1\over 2} \right)^2
+f^s _{eq} (\rho)
\(generalfbeta)
$$
which is clearly a straightforward generalization
of  \(newone).

The definition of $\sigma_s$ for a system with short range interactions in
the one phase system 
can be found in Part II of [\rcite{Spohn}],
 formulas (2.27), (2.71) and (2.72) (in our notation
formulas (2.27) and (2.71) must be multiplied by $\beta$).
It is given by a Green-Kubo formula and is related to the diffusion
coefficient of the reference system $D_s(\rho)$ by the relation,
$$
\sigma_s = \chi_s D_s
\(yaels)
$$
where $\chi_s(\rho)$ us the inverse of the derivative of the chemical
potential, i.e.\ the compressibility of our reference system is defined as
$$
\chi_s \equiv  {1\over {f^s_{eq}}^{\prime \prime} (\rho)}=
\beta \sum_{x \in {\Bbb Z}^d}
\left(
\int \eta (x) \eta (0) \text{d} \mu^s_\rho (\eta) -\rho^2
\right)
\(compressibility)
$$

While a complete proof of \(generalfbeta) for general short range
interactions is lacking at the moment, some particular cases can be handled
[1].  
In order to justify \(generaleq) on a heuristic level we will draw
an analogy with Section 3.4 of Part II of [{\rcite{Spohn}].
There a linear response argument is developed
for the system with Hamiltonian
$$
H_V(\eta)
=H_s(\eta)+\sum_{x \in \Lambda _\gamma}
V(\epsilon x) \eta (x)
\(drivenlattice)
$$
where $V(r)$ is a smooth function from $T^d$ to
${\Bbb R}$ and $\epsilon>0$ is a small
parameter. This is the Hamiltonian of a weakly driven
lattice gas: the dynamics is defined by the Poisson
rates $c_{\epsilon,V} (x,y; \eta)=
\Phi (\beta \Delta_{x,y} H_V (\eta))$ for $\vert x-y \vert=1$
(and $c_{\epsilon,V}(x,y;\eta)=0$ otherwise).
Hence we have
$$
c_{\epsilon,V}(x,x+e; \eta)=
\(secondobserve)
$$
$$
\Phi(\beta \Delta_{x,x+e} H_s(\eta))
-{\beta \epsilon}\Phi^\prime(\beta\Delta_{x,x+e} H_s(\eta))
(\eta (x+e)-\eta (x) )
(e\cdot \nabla V) (\epsilon x)+O(\epsilon ^2)
$$

In the hydrodynamical scaling limit, with  $r$ scaled as  $\epsilon x$,
$\epsilon \to 0$,    external force will vanish like
$\epsilon$.  Hence it can be argued, see 
 in  the macroscopic continuity equation for the density
$\partial \rho / \partial t = -\nabla \cdot J$, the current
$J$ should be the sum of a diffusive term, unaffected by the weak external
force, and a drift term proportional to $F = -\nabla V(r)$, 


$$
J= -D_s(\rho) \nabla \rho + \sigma(\rho)F
\(current)
$$
with $\sigma(\rho)$ a mobility to be determined.  To obtain $\sigma$ 
 we observe that the 
Gibbs measure
$\mu^{\beta}_V $ associated with $H_V(\eta)$ at the temperature
$1/\beta$
 is a stationary state of
our system.  This implies that ${\partial \rho (r,t)
 / \partial t} = 0$ whenever
$\rho(r,0) = \rho_{eq}(r)$ is the density obtained from $\mu^{\beta}_V$.  In
the limit $\epsilon \to 0$, $\rho_{eq}$ is the density which minimizes
the functional [\rcite{LP}]
$$
{\Cal F}_V (\rho) = \int \left[f^s _{eq} (\rho(r)) + V(r) \rho(r) \right]
\text{d}r
\(drivenfreeen)
$$
 with a total mass constraint. In particular,
at the minimum,
$\delta{\Cal F}_V /\delta \rho = const$.  But, according to
\(current),
 $\rho({
r}, t)$ satisfies the equation
$$
{\partial \rho \over \partial t} = \nabla \cdot[D_s(\rho) \nabla \rho +
 \sigma (\rho) \nabla V]
\(guess)
$$
so the requirement that $\partial \rho / \partial t = 0$ at $\rho({ r}) =
\rho_{eq}({ r})$
then requires that $\sigma =D_s \chi$.  This gives 
the described Einstein relation,
showing that $\sigma = \sigma_s$.

In the model we are considering $\epsilon =\gamma$ and
 the external potential $V$ is replaced by
an internally generated one, $ -\gamma^d \sum_{y \in \Lambda_{\gamma}}
J(\gamma(x-y))\eta(y)$.  Taking now the hydrodynamic scaling limit, with
the scaling parameter $\epsilon$ set equal to $\gamma$ and letting $\gamma
\to 0$,  the previous
analyses should remain valid with $V (r) \to -\int J(r -
r^\prime)\rho(r^\prime)dr^\prime$ in \(guess) which then gives 
\(generaleq).  

We note that if we look at a hydrodynamical space 
scale $\epsilon^{-1}$ which is
much larger than $\gamma^{-1}$, say $\epsilon = \gamma^\delta$, $\delta > 1$,
then we get a purely diffusive equation for $\rho(r,t)$.  This is obtained
from \(generaleq) by dropping the second term in \(newway).  This equation
holds rigorously {\it outside} the spinodal region,
 where the new diffusion
constant becomes negative, see [\rcite{LOP}] and [\rcite{G1}].

\head
4. A remark on the effect of an external driving force
\endhead
\resetall
\sectno=4

As one may expect, domain coarsening phenomena
become richer and more complicated in the presence
of an external driving force
and some important points are still controversial
(see e.g. [\rcite{Laberge},\rcite{YMHJ}]). One of the interesting
problems is understanding the scaling behavior
of the cluster size;  different behaviors
are expected depending on whether the characteristic length  is measured
in the direction perpendicular or parallel to the field.

Here we simply observe that it is straightforward to extend the
results of this paper to the case of a system on which a weak external
force is acting. This system has a Hamiltonian
$$
H(\eta)+\sum_{x \in \Lambda _\gamma}
V(\gamma x) \eta (x)
\(externalforce)
$$
where $V$ is a smooth function from $T^d $ to $\real$, as in Section 3.
Remember that $H=H_s+H_\gamma$ (formula \(HL)).
The discussion in Section 3 easily gives the following guess
for the hydrodynamic limit of this system
$$
{\partial \rho \over \partial t} =
\nabla \cdot\left[
\sigma \nabla
\left(
{\delta {\Cal F} \over \delta \rho }+V
\right)
\right]
\(generaleqdriven)
$$
alternatively one can redefine ${\Cal F}(\rho)$
as  ${\Cal F}(\rho)-\int V(r) \rho(r) \text{d}r$ and the evolution
equation would still be \(generaleq). We observe
that the proofs in Section 4 directly extend to
this case, if $K=0$.

The situation is similar in the case of an external force
which is not the gradient of a potential $V$, e.g.\
a constant force $E\in \R^d$. In this case
the rates of the process would be
$$
\Phi\left(\beta (H(\eta^{x,y})-H(\eta)
-\gamma E\cdot (x-y)(\eta(x)-\eta(y))
\right)
\(constantext)
$$
if $\vert x-y \vert=1$ and zero otherwise  and
the macroscopic equation
$$
{\partial \rho \over \partial t} =
\nabla \cdot\left[
\sigma \nabla
\left(
{\delta {\Cal F} \over \delta \rho }
\right)
\right]+\nabla \cdot \left(E
\sigma\right).
\(constantexteq)
$$

The term added  to \(generaleq) to obtain
in \(constantexteq) is exactly the term
which in [\rcite{YMHJ}] is added to the C--H
equation to adapt it to a situation in which
an external field is present. Problems and limitations
to the use of macroscopic models like
\(constantexteq) are discussed in [\rcite{Laberge}]:
\(constantexteq) in fact does not seem to be suitable
to describe systems with strong ($O(1)$ and not, like here,
$O(\gamma)$) external fields.

\head
5. The hydrodynamic limit of the particle system: proofs
\endhead
\resetall
\sectno=5

The Proof of Theorem 1 is an immediate consequence of
Theorem 2 and Proposition 1 given below.

\definition{Some preliminary definitions:}
In this section we will denote by $e\in \Z^d$ a unit lattice vector.
By $\sum _e$ we will mean the sum over all $e_j\in \Z^d$,
$(e_j)_i=\delta _{j,i}$ ($i,j \in \{1, \ldots ,d\}$).
Let us recall the Poisson rates
$$
c_\gamma (x,x+e;\eta)=\Phi(\beta \Delta_{x,x+e} H_\gamma(\eta))
\ \  \ \ \ \ x \in \Lambda_\gamma , \ \eta \in \Omega_\gamma
\(allrates)
$$
where we used the notation
$$
\Delta_{x,y} H_\gamma(\eta) =
H_\gamma(\eta^{x,y})-H_\gamma(\eta)
$$
for $x,y \in \Lambda_\gamma$.
We recall that $\Phi \in C^2$ satisfies \(detbalance),
$\Phi(0)=1$,
and that $c_\gamma (x,y; \eta)=0$ if $\vert x -y \vert \not=1$.
In the proofs we shall denote by $c_\gamma^0(x, y;\eta)$ the rates
in the case in which $J\equiv 0$, that is
$$
c_\gamma^0 (x,x+e;\eta)=1 \ \ \ \ \ \ \ \ \ \
 x \in \Lambda_\gamma
\(recallrates)
$$
This process is called {\sl the simple exclusion process}: the process
with $J\not=0$ will be treated as a perturbation of
the simple exclusion process. Moreover we will denote by
$c_\gamma ^{p} (x, y; \eta)$ the rates when
$\Phi(E)=\exp(-E/2)$ ($E\in \real$).
The law of the simple exclusion process with initial condition $\mu$
is denoted by $\Prob ^{0, \mu}_\gamma$ ($\E^{0, \mu}_\gamma$), while
the law of the process with rates $c_\gamma ^p(x,y; \eta)$
is denoted by $\Prob ^{p, \mu}_\gamma$ ($\E^{p, \mu}_\gamma$).

Let $M_1$ be the set of measurable functions $\rho :T^d\rightarrow
[0,1]$. $M_1$ is equipped with the weak topology induced by duality
by $C(T^d)$, the continuous real--valued functions
on $T^d$, according to
$$
\langle
\rho , G
\rangle =\int_{T^d} G(r) \rho(r) \text{d}r
$$
where $\rho \in M_1$ and $G\in C(T^d)$.
Given $\gamma>0$ we define the empirical measure of the process at $r$ as
$$
\rho_\gamma (r; \eta)=
\sum_{x\in \Lambda_\gamma} \eta (x) \chi_{\prod_{\ell =1} ^d
[x_i \gamma, (x_i +1) \gamma ]}
(r)
\(rhogamma)
$$
where
 $\chi_A$ denotes the characteristic function of the set $A
\subset T^d$.
The empirical averages of a function $f:\Omega _\gamma \rightarrow \real$
over a ball of radius $R>0$
is defined as
$$
\text{Av}_{R,x} f =
{1\over \vert B(R) \cap \Z^d \vert}
\sum_{y \in B(R)\cap \Z^d }
f(\tau _{x+y} \eta)
\(empaverage)
$$
where $B(R)=\{r : \vert r \vert \le R \}$,
$x\in \Lambda_\gamma$ and $\tau_z:
\Omega_\gamma \rightarrow
\Omega_\gamma$ is defined by
$(\tau_z \eta)(x)=\eta(x+z)$.  For $\alpha \in [0,1]$,
let $\nu_{\alpha}^B$ be the Bernoulli measure over $\Omega _\gamma$
 such that $\nu_\alpha ^B (\eta (x))=\alpha$ for all $x\in \Lambda_\gamma$.
We set
$$
\tilde f (\alpha )
= \nu_\alpha ^B (f).
\(localav)
$$

Given $\eta \in D([0, \infty), \Omega_\gamma )$,
the space of right continuous functions from $[0, \infty)$ to
$\Omega_\gamma$ with limit from the left,
equipped with the Skorohod topology,
for every $t\in \real ^+$ and $r \in T^d$
we define
$$
\rho_\gamma(r,t)= \rho_\gamma (r;\eta _{\gamma^{-2}t})
\(rhogammat)
$$
We will adopt the notation $\rho_\gamma =\rho_\gamma (\cdot, \cdot)$
and when we want to keep the time fixed we will write
$\rho_{\gamma,t}$, which stands for $\rho_\gamma (\cdot ;\eta
_{\gamma^{-2}t})$. So
$\rho_{\gamma,t} \in M_1$ and $\rho_\gamma \in D([0, \infty); M_1)$.

With this notation,
 the family of measures $\{\mu _\gamma \}_{\gamma
>0}$ is an initial condition associated with $\rho_0$ if for
any $\delta >0$ and
any $\phi \in C (T^d)$ we have
that
$$
\lim _{\gamma \rightarrow 0}
\E^{ \mu_\gamma}_\gamma \left( \left \vert
\langle \rho_{\gamma, 0}, \phi\rangle-
\langle \rho_0 , \phi \rangle
\right\vert >\delta
\right)=0.
\(initial)
$$
(compare with \(initprob)).

A function $\rho \in  C([0,\infty), M_1)$ is a
weak solution of
\(maineq)  if
for all $t \in \real^+$ and all $\phi \in C^{2,1} (T^d
\times [0, \infty))$,
$\rho $ satisfies
$$
\int_{T^d} \rho (r,t) \phi (r,t) \text{d} r -
\int _{T^d} \rho_0 (r ) \phi (r,0) \text{d} r =
$$
$$
\int \int _{Q_t}
\rho (r, s) \partial_s \phi (r,s)
\text{d} r \text{d} s
+\int \int_{Q_t}
\rho  (r,s) \Delta \phi (r,s)
\text{d} r \text{d} s +
\(weakform)
$$
$$
\beta \int \int _{Q_t}
\rho (r,s)(1-\rho (r,s))
(\nabla J * \rho)(r,s) \nabla \phi (r,s)
\text{d} r \text{d} s
$$
in which
$Q_{t}=T^d\times [0, t)$.

Finally, for $f\in C^k$ ($k\in \Z^+$), we  set
$\Vert f \Vert _{C^k} = \Vert f \Vert _{C^0}+ \sum_{i_1,\ldots, i_d}
\Vert \partial_{i_1} \ldots \partial_{i_d} f \Vert_{C^0}$ ($i_1,
\ldots , i_d \in \Z^+$ and
$\vert i_1+\ldots + i_d\vert =k$), where
$\Vert \cdot \Vert_{C^0}$ is the sup--norm. $\Vert \cdot \Vert_p$ will
denote the $L^p$--norm.

\enddefinition

\proclaim{Theorem 2}
For any $t>0$
$$
\lim _{\gamma \rightarrow 0}
\Prob^{\mu_\gamma}_\gamma \left( \left \vert
\langle \rho_{\gamma, t}, \phi\rangle-
\langle \rho_t , \phi \rangle
\right\vert >\delta
\right)=0.
\(hydrolim)
$$
where $\rho_t$ is the unique solution of \(weakform).
\endproclaim

\vskip 0.3 cm
{\it Remark:} a byproduct of the proof of Theorem 2 is  that
the random variable
$\rho_\gamma\in D([0,\infty); M_1)$ converges weakly, as $\gamma$
approaches zero, to
the deterministic trajectory $\rho\in C([0,\infty); M_1)$,
unique solution of \(weakform).

\vskip 0.2 cm

{\it Proof of Theorem 2:}
For all $\epsilon >0$, $f: \Omega _\gamma \rightarrow \real$ and $\eta \in
\Omega _\gamma$  set
$$
R_{\epsilon , \gamma}
(f; \eta)
=
\sum_x \left\vert
\text{Av} _{\epsilon \gamma ^{-1},x}
(f;\eta) - \tilde f (\text{Av}_{\epsilon \gamma ^{-1},x}
(\pi_0;\eta)))\right\vert
\(Rempav)
$$
in which $\pi_0: \Omega _\gamma \rightarrow \{0,1\}$
is defined as $\pi_0(\eta)=\eta (0)$.

Let us recall the following result which is a straightforward extension
to any dimension of Proposition 2.1 of [\rcite{KOV}].

\proclaim{Lemma 1}
Given a cylindrical\footnote{
A  function $f$ on $\{0 ,+1\}^{\Z^d}$ is said to be
cylindrical if there exists a finite set $A\subset \Z^d$
such that $f(\eta)=f(\eta ^\prime)$ whenever $\eta(x)=\eta ^\prime (x)$
for all $x \in A$. Hence any cylindrical function has an obvious
 restriction to $\Omega_\gamma$, provided that $\gamma$ is small
enough (i.e. if  $A \subset \Lambda_\gamma$). The $f$ which appears
in formula \(Lemma1) is precisely this restriction.}
 function $f$, $t \in \real ^+$ and $\delta >0$
$$
\limsup_{\epsilon \rightarrow 0}
\limsup_{\gamma \rightarrow 0}
\gamma ^d \log
\Prob_\gamma ^{0,\nu^B_{1/2}}
\left( \gamma ^d \int _0 ^{\gamma^{-2}t}
R_{\epsilon, \gamma} (f ; \eta _s) \text{d}s \ge \delta \right)=
-\infty
\(Lemma1)
$$
\endproclaim

We have to extend Lemma 1
to the case of $J\not=0$ and general initial condition.
We are not going to extend the Lemma to the general case,
but only to $\Prob_\gamma ^{p, \mu_\gamma}$: in the general case we will
obtain a weaker, but sufficient, estimate via a relative entropy
argument.
In order to extend Lemma 1, it is
 sufficient to show that there is
a constant $c$ such that
$$
\gamma ^d \log
\left(
{
\text{d} \Prob_\gamma ^{p,\mu_\gamma}
\over
\text{d}
\Prob_\gamma ^{0,\nu^B_{1/2}} }(\{ \eta _t\} _{t \in [0, \gamma^{-2}\tau]})
\right)
\le c
\(ctrlRN)
$$
for all $\eta \in D([0, \gamma^{-2}\tau]; \Omega_\gamma)$,
since \(Lemma1) and \(ctrlRN)
easily imply
$$
\limsup_{\epsilon \rightarrow 0}
\limsup_{\gamma \rightarrow 0}
\gamma ^d \log
\Prob_\gamma ^{p, \mu_\gamma}
\left( \gamma ^d \int _0 ^{\gamma^{-2}t}
R_{\epsilon, \gamma} (f , \eta _s) \ge \delta \right)=
-\infty
\(superexp)
$$
under the same conditions stated in Lemma 1.
The bound \(superexp) will be obtained
 via direct evaluation of the Radon--Nikodym  derivative, whose explicit
expression is the following
$$
\log\left(
{\text{d}\Prob_\gamma^{p, \mu_\gamma}
\over
\text{d}\Prob_\gamma^{0, \nu^B_{1/2}}
}(\{\eta _t \}_{t \in [0, \gamma^{-2}\tau]} ) \right)
=
$$
$$
\log \left(
{\text{d} \mu_\gamma \over \text{d} \nu^B_{1/2} }
(\eta _0) \right)-
\(RN0)
$$
$$
\int_0^{\gamma^{-2}\tau} \sum_{x,e}
\left[c_\gamma^{p}(x,x+e;\eta_t)-
c_\gamma^{0}(x,x+e;\eta_t)\right] \text{d} t+
\(RN1)
$$
$$
\sum _{x,e}
\int_0^{\gamma^{-2}\tau}
\log\left({
c_\gamma^{p}(x,x+e;\eta_{t^-})\over
c_\gamma^{0}(x,x+e;\eta_{t^-})}
\right)
\text{d}  {\Cal J}_t^{x,x+e}
\(RN2)
$$
in which ${\Cal J}_t^{x,x+e}$ is the Poisson process that counts
the exchanges of occupation number between
 $x$ to $x+e$ in the time
interval $[0,t]$.
We will bound the three terms of the RN derivative
(\(RN0), \(RN1) and  \(RN2) )
separately. Let us observe that for \(RN0), since
$(\text{d}\mu_\gamma / \text{d} \nu^{B}_{1/2})
(\eta) \le 2^{\gamma^{-d}}$,
the desired estimate is immediately obtained.

Now and later on we will make use of the following Lemma.

\proclaim{Lemma 2}
For $x\in T^d$ and $\eta \in \Omega _\gamma$, let
$$
h^{x,e}_\gamma (\eta)=
\Delta_{x,x+e} H_\gamma(\eta)-{\gamma}
 (\eta(x+e)-\eta(x))
\left[ \gamma^{d}\sum_z \eta (z) (e \cdot\nabla
 J)(\gamma (x-z))\right]
\(hgamma)
$$
There exists a constant $c_1$ such that for every  $x \in
\Lambda _\gamma$, every $\eta \in
\Omega_\gamma$ and every $e$
$$
\vert h^{x,e}_\gamma (\eta) \vert \le c_1 \gamma ^2
\(deltaE)
$$
in particular there is a   constant $c_2$ such that for all $x \in
\Lambda _\gamma$ and all
$\eta\in \Omega_\gamma$
$$
\vert \Delta_{x,x+e} H_\gamma(\eta)\vert \le
c_2 \gamma.
\(deltaErough)
$$
\endproclaim

{\it Remark:} an immediate consequence
of Lemma 2 is that the rates are bounded: for any $J$, $\beta$ and $\Phi$,
there exists $\gamma_0$ such that
$$
\sup_{\gamma \in (0, \gamma_0)}\sup _{\eta , x,y }
c_\gamma (x,y; \eta) \le 2 .
\(ratesarebounded)
$$

{\it Proof of Lemma 2:}
By definition
$$
-2
\Delta_{x,y}H_\gamma(\eta)=
\sum_{z,z^\prime}
[\eta^{x,y}(z) \eta^{x,y}(z^\prime)-
\eta(z)\eta(z^\prime)]\gamma ^d J((z-z^\prime)\gamma)
\(fty0)
$$
add and subtract in the square brackets
$\eta^{x,y} (z) \eta (z^\prime )$ and use the symmetry $J(r)=J(-r)$
to get
$$
\align
-2
\Delta_{x,y}H_\gamma(\eta)&=
(\eta(y)-\eta(x))\sum_z
(\eta^{x,y}(z)+\eta (z))
\gamma^d[J(\gamma(z-x))-J(\gamma(z-y))]
\\
&=-2
(\eta(y)-\eta(x))\sum_z
\eta (z)
\gamma^d[J(\gamma(y-z))-J(\gamma(x-z))]
\(fty1)
\\
&\ + 2(\eta(y)-\eta(x))^2\gamma ^d[J(0)-J(\gamma(x-y))].
\(fty2)
\endalign
$$
and if $\vert x-y \vert =1$
the modulus of the term \(fty2) can be bounded by $2 \Vert
J\Vert_{C_1}\gamma ^{d+1}$
and we can
substitute the discrete gradient in
 \(fty1) with $(x-y)\cdot\nabla$ and the error
will be  bounded by $\Vert J \Vert_{C^2} \gamma^2$, so that \(deltaE)
is proven with $c_1= 2 \Vert J\Vert_{C_1}+\Vert J \Vert_{C^2}$.
Formula \(deltaErough) follows
immediately from \(hgamma) and \(deltaE).
This ends the proof of Lemma 2.\qed

Let us go on with the proof of Theorem 2.
The term in \(RN1) can be written as
$$
\int _0 ^{\gamma^{-2}\tau} \sum_{x,e} \eta_t(x) (1-\eta _t(x+e))
\left[
c_\gamma ^p (x, x+e; \eta _t)-1
\right]
\text{d} t
\(RN2.1)
$$
By Lemma 2 we obtain that
\(RN2.1) is equal to
$$
  \int_0^{\gamma^{-2}\tau} {1\over 2}
\sum_{x,e}
\bigg\{
\eta_t(x)(1-\eta_t (x+e))
$$
$$
\times
\left[
\exp\left(
{\beta \gamma \over 2}
\gamma^{d}
\sum _z \eta _t(z) (e \cdot\nabla J)(\gamma (z-x)) +
{\beta\over 2} h_\gamma^{x,e}(\eta_t)\right)
-1 \right]
+\eta_t(x+e)(1-\eta_t (x))
$$
$$
\times
\left[
\exp\left(-
{\beta \over 2}
\gamma^{d+1}
\sum _z \eta _t(z) (e \cdot\nabla J)(\gamma (z-x-e))
+{\beta\over 2} h_\gamma^{x+e,-e}(\eta_t)
\right)
-1 \right]
\bigg\}
\text{d} t  \(pyt)
$$
Set $g_\gamma (x;\eta)=(\beta/2)(\gamma ^d \sum_z \eta (z) (e \cdot\nabla
J)(\gamma (z-x))$.
The modulus of \(pyt) can be bounded by
$$
\align
\bigg\vert
 \int _0 ^{\gamma ^{-2}\tau} {1\over 2}
\sum_{x,e} \bigg\{
&-\eta _t (x)\eta _t (x+e) \left[
\exp(\gamma g_\gamma (x;\eta_t))+
\exp(-\gamma g_\gamma (x+e;\eta_t))-2 \right]
\\
&+\eta _t (x) \left[
\exp(\gamma g_\gamma (x;\eta_t))+
\exp(-\gamma g_\gamma (x;\eta_t))\right]
\bigg\} \text{d} t \bigg\vert
\\
&
+2 (c_1 \gamma^2) (\tau \gamma^{-2}) (d \gamma^{-d}) \le c \gamma^{-d}
\endalign
$$
in which we used
\(ratesarebounded),
\(deltaE), \(deltaErough),
the fact that $\vert g_\gamma (x; \eta)-g_\gamma (x+e; \eta )\vert \le
(\beta /2)\Vert J\Vert _{C^2} \gamma$ and
$c=c(d, \tau, \beta, \Vert J \Vert_{C^2})< \infty$.
The third and last piece \(RN2) in the RN derivative is treated as
follows. Given a sample path $\{\eta_t:t \in [0,\gamma ^{-2}\tau]\}$,
call $n=\sum _{x,e} {\Cal J}_{\gamma^{-2}\tau} ^{x,x+e}$ the total number of
exchanges in the interval $[0,\gamma^{-2} \tau]$, which is finite for every
$\eta_\cdot
\in D([0,\gamma^{-2}\tau], M_1)$.
The $i^{\text{th}}$ jump $(i=1,\dots ,n)$ happens at $t=t^i$
and it moves a particle from $x^i$ to $x^i+e^i$.
The last term in the RN derivative is equal to
$$
\align
\sum_{x,e} \int_0^{\gamma ^{-2}\tau} \log
\left({c_\gamma ^p (x,x+e;\eta_{t^-}) \over c_\gamma ^0(x,x+e;\eta_{t^-})}
\right) \text{d} {\Cal J}_t ^{x,x+e}
=&
\sum_{i=1}^n (\Delta _{x^i, x^i+e^i}H_\gamma)(\eta_{t_i^-})
\\
=& H_\gamma(\eta _{t \gamma^{-2}})-H_\gamma(\eta_0)
\(RN3.1)
\endalign
$$
which is easily bounded
by $2 \max_{\eta} \vert H_\gamma (\eta ) \vert \le (\Vert
J \Vert _{C^0}) \gamma ^{-d}$ and so the proof of \(ctrlRN)
(and so of
\(superexp)) is complete.

Clearly \(superexp) implies that for  $\epsilon$ sufficiently small
$$
\lim_{\gamma \rightarrow 0}
\Prob_\gamma ^{p, \mu_\gamma}
\left( \gamma ^d \int _0 ^{\gamma^{-2}t}
R_{\epsilon, \gamma} (f , \eta _s) \ge \delta \right)=0
\(spxP)
$$
The aim is to obtain a similar inequality for
 $\Prob_\gamma ^{ \mu_\gamma}$.
To this purpose let us introduce the relative entropy between
$\Prob_\gamma ^{ \mu_\gamma}$ and $\Prob_\gamma ^{p, \mu_\gamma}$
on the time interval $[0, \gamma^{-2}\tau]$:
$$
H^\gamma_\tau\left( \Prob_\gamma ^{ \mu_\gamma} \big\vert
\Prob_\gamma ^{p, \mu_\gamma}\right)=
\E_\gamma ^{ \mu_\gamma} \left[
\log
\left(
{\text{d}\Prob_\gamma^{ \mu_\gamma}
\over
\text{d}\Prob_\gamma^{p, \mu_\gamma}
}(\{\eta _t \}_{t \in [0, \gamma^{-2}\tau]} ) \right)
\right]
\(relentropy)
$$
and we will evaluate \(relentropy) directly by writing
the RN derivative explicitely. The obtained expression
is absolutely analogous to \(RN1) and \(RN2) (now the expressions
 $c_\gamma- c_\gamma^p$ and $c_\gamma/ c_\gamma^p$ are replacing
$c_\gamma^p-c_\gamma ^0$ and $c_\gamma^p/c_\gamma ^0$  respectively)
and there is no term \(RN0), since the initial condition is the same.
Observe that by differentiating \(detbalance)
we obtain
$$
\Phi ^\prime(E)= -\exp(-E) \Phi(-E) -\Phi ^\prime(E)\exp(-E)
\(diffdetbal)
$$
and so $\Phi^\prime (0)=-\Phi(0)/2=-1/2$.
This implies that there exists a
constant $c$ such that $\vert c_\gamma (x,y;\eta)-
c_\gamma ^p (x,y;\eta)\vert \le c\gamma^2$ and
$\vert \log( c_\gamma (x,y;\eta)/
c_\gamma ^p (x,y;\eta))\vert \le c\gamma^2$.
Hence
$$
H^\gamma_\tau\left( \Prob_\gamma ^{ \mu_\gamma} \big\vert
\Prob_\gamma ^{p, \mu_\gamma}\right)
\le (d \gamma ^{-d}) (\gamma ^{-2} \tau) (c \gamma ^2)+
 (c \gamma ^2)\E^{\mu_\gamma}_\gamma (n) \le c^\prime \gamma ^{-d}
\(entest)
$$
in which $n$ is the total number of Poisson exchanges in
$[0,\gamma^{-2} \tau]$. The bound in \(entest) follows directly from
\(ratesarebounded).
By applying a well known entropy inequality (see e.g.
[\rcite{Yauentropy}], formula (2.18)) to our set up
we obtain
$$
\Prob_\gamma ^{\mu_\gamma}
\left( \gamma ^d \int _0 ^{\gamma^{-2}t}
R_{\epsilon, \gamma} (f , \eta _s) \ge \delta \right)
\le
{ \log 2 + H^\gamma_\tau\left( \Prob_\gamma ^{ \mu_\gamma} \big\vert
\Prob_\gamma ^{p, \mu_\gamma}\right)
\over \log \left[
1+1/
\Prob_\gamma ^{p, \mu_\gamma}
\left( \gamma ^d \int _0 ^{\gamma^{-2}t}
R_{\epsilon, \gamma} (f , \eta _s) \ge \delta \right)
\right]}
\(yauineq)
$$  From \(superexp), \(entest) and \(yauineq) directly
follows that
$$
\lim_{\eps \rightarrow 0}
\limsup_{\gamma \rightarrow 0}
\Prob_\gamma ^{\mu_\gamma}
\left( \gamma ^d \int _0 ^{\gamma^{-2}t}
R_{\epsilon, \gamma} (f , \eta _s) \ge \delta \right)=0
\(generalbound)
$$

Let $\phi \in C^{2,1}({T^d}\times [0, \infty))$:
we will use the notation $\phi_t (\cdot)=
\phi(\cdot, t)$. Define
$$
\langle \rho_{\gamma ,t}, \phi_t \rangle_\gamma=
\gamma ^d
\sum _x \eta_{\gamma^{-2}t} (x) \phi(\gamma x ,t)
\(easeofreader)
$$
and observe that for any $t>0$ there exists $c=c(\phi)$ such that
$$
\left\vert
\langle \rho_{\gamma ,t}, \phi_t \rangle_\gamma
-
\langle \rho_{\gamma ,t}, \phi_t \rangle\right\vert \le
c\gamma
\(smallcorrection)
$$
which makes the two quantities appearing in the left--hand side
of \(smallcorrection) interchangeable for our purposes.
Let us define
$$
\Gamma _{1, \gamma }(t)=\gamma ^{-2}
L_\gamma \left (
\langle
\rho_{\gamma ,t}, \phi_t\rangle_\gamma \right)+
{\partial _t}\left (
\langle
\rho_{\gamma ,t}, \phi_t\rangle_\gamma \right)
\(gammauno)
$$
where with partial derivative in $t$ we mean the derivative
with respect to the time dependence of $\phi$,
and
$$
\Gamma _{2, \gamma }(t)=\gamma ^{-2}
L_\gamma \left[\left (
\langle
\rho_{\gamma ,t}, \phi_t\rangle_\gamma \right)^2
\right]-
2 \left (
\langle
\rho_{\gamma ,t}, \phi_t\rangle_\gamma \right)
L_\gamma
\left (
\langle
\rho_{\gamma ,t}, \phi_t\rangle_\gamma \right).
\(gammadue)
$$
It is well known (see for instance [\rcite{DP}]) that
$$
M_{\gamma }(\tau; \phi)
\equiv
\langle
\rho_{\gamma ,\tau}, \phi_\tau\rangle_\gamma -
\langle
\rho_{\gamma ,0}, \phi_0\rangle_\gamma -
\int_0^\tau \Gamma _{1, \gamma }(t) \text{d}t
\(mart1)
$$
and
$$
N_{\gamma }(\tau; \phi) \equiv
\left(M_{\gamma }(\tau; \phi)\right)^2
-\int_0^\tau \Gamma _{2, \gamma }(t) \text{d}t \(mart2)
$$
are $\Prob _\gamma ^{\mu_\gamma}-$martingales with respect to the
filtration generated by the family of random variables
 $\{\eta_\tau \}_{\tau \in \real ^+}$. Moreover
they vanish at $\tau =0$.

Let us start by computing $\Gamma_{1, \gamma}$ and $\Gamma_{2,
\gamma}$.
By using the definition of the generator,
we obtain
$$
\align
\Gamma _{1, \gamma }(t)=&\\
\gamma ^{-2}
\gamma ^d \sum_{x,e} c_{ \gamma}
(x,x+e;\eta _{\gamma^{-2}t})&
\left[(\phi (\gamma x, t)-\phi (\gamma (x+e), t)
(\eta_{\gamma^{-2}t}(x+e)-\eta_{\gamma^{-2}t}(x))\right]\\
&\ \ \ \ \ \ \ \ \ +\gamma ^d \sum_x \partial_t
\phi(\gamma x ,t)\eta _{\gamma^{-2}t}(x)
\(quasilapl)
\endalign
$$
and the first term on the right--hand side of
\(quasilapl) can be rewritten
as
$$
\gamma ^d \sum_x (\Delta \phi)(\gamma x ,t) \eta
_{\gamma^{-2}t}(x)+ \(oralapl)
$$
$$
 \gamma ^d \sum_{x,e}
\eta_{\gamma^{-2}t}(x)(1-\eta_{\gamma^{-2}t} (x+e))\bigg[{\beta\gamma ^d
\over 2}
 \sum _z \eta _{\gamma^{-2}t}
(z) (e \cdot\nabla J)(\gamma (x-z)) \bigg]
(e \cdot\nabla\phi) (\gamma x, t)
$$
apart for corrections that are uniformly vanishing as $\gamma$ goes to
zero
and so $\vert \Gamma _{1, \gamma} (t)\vert$ is uniformly bounded
(this follows from the observation following \(diffdetbal) that
$\Phi^\prime (0)=-1/2$).
Moreover
$$
\align
&=
\Gamma _{2, \gamma }(t)\\
&=\gamma ^{-2}
 \sum_{x,e} c_{\gamma}
(x,x+e;\eta _{\gamma^{-2}t}) \left[
\gamma^d \sum_z \eta_{\gamma^{-2}t} ^{x,x+e}(z) \phi(z,t)-
\gamma^d \sum_z \eta_{\gamma^{-2}t} (z) \phi(z,t)
\right]^2
\\
&=\gamma^{-2} \gamma^{2d}
\sum _{x,e} c_{\gamma} (x, x+e;\eta _{\gamma^{-2}t})
\left[ \phi(\gamma (x+e),t)-\phi(\gamma x,t)
\right]^2
\(g2is)
\endalign
$$
and from \(g2is) we obtain
$$
\vert \Gamma _{2, \gamma }(t)\vert
\le 2 \Vert \phi \Vert _{C^1} \gamma ^d .\(g2vanishes)
$$
The fact that $\vert \langle \rho_{\gamma, t}, \phi_t\rangle \vert$,
$\vert \Gamma_{1, \gamma }(t)\vert$ and $\vert \Gamma_{2, \gamma
}(t)\vert$
are bounded uniformly in $t \in [0,\tau]$ ($\tau \in \real ^+$), $\eta
\in D([0, \tau];\Omega _\gamma)$ and $\gamma>0$,
immediately implies that the family of
random variables $\{\rho_\gamma\}_{\gamma>0}$ is (sequentially) relatively
compact in $D([0, \infty);M_1)$ (see for instance
[\rcite{KOV}] and [\rcite{GPV}]).
Moreover every limit point $\rho$ of $\{\rho_\gamma\}_{\gamma>0}$
lies in $C([0, \infty);M_1)$, which follows directly
from the fact that with $\Prob _{\gamma}^{\mu_\gamma}-$probability
one the $\sup_t \vert \sum_{x,e} ({\Cal J}_t ^{x,x+e}-
{\Cal J}_{t^-} ^{x,x+e})\vert =1$ and so the quantity in
\(easeofreader) can have jumps of size at most
$\Vert \phi \Vert_{C^0} \gamma$, which vanish in the limit
(see [\rcite{DP}], Theorem 2.7.8). We are then left with
the task of identifying the limit.

Let us observe that \(g2vanishes) and  Doob's
martingale inequality imply that
for every $\delta_1 >0$, $\phi \in C^{2,1}$ and $\tau \in \real^+$
$$
\lim_{\gamma \rightarrow 0}
\Prob _\gamma ^{\mu_\gamma}\left(
\sup_{t\in [0, \tau]}
\vert M_\gamma (t;\phi)
\vert >\delta_1
\right) =0. \(g2anddoob)
$$
Observe that, because of the smoothness of $\phi $ and  $J$,
for every $\delta _2>0$, there is an $\epsilon _0>0$ such that for all
$\epsilon\in (0, \epsilon_0]$ the following holds
$$
\bigg\vert
\beta \int_0^\tau
\gamma ^d \sum _{x,e} \pi_x(\eta _t)
\pi_{x+e}(\eta _t)
(\rho _{\gamma ,t}*e \cdot\nabla J)(\gamma x)
(e \cdot\nabla \phi) (\gamma x,t) \text{d}t
$$
$$
- \int_0 ^\tau \gamma ^d \sum _{x,e}
\text{Av}_{\epsilon^{-1}\gamma, x}
(\pi_0 \pi_e, \eta _t)  \beta
(\rho _{\gamma ,t}*e \cdot\nabla J)(\gamma x)
(e \cdot\nabla \phi)
\text{d}t
\bigg\vert \le \delta_2
$$
and by
\(generalbound)
we obtain that for every $\delta>0$
$$
\lim_{\epsilon \rightarrow 0}
\limsup_{\gamma \rightarrow 0}
\Prob_\gamma ^{\mu_\gamma}
\bigg(
\bigg\vert
\beta \int_0^\tau
\gamma ^d \sum _{x,e}
\text{Av}_{\epsilon^{-1}\gamma, x}
(\pi_0 \pi_e, \eta _t)
(\rho _{\gamma ,t}*e \cdot\nabla J)(\gamma x)
(e \cdot\nabla \phi) (\gamma x,t) \text{d}t
$$
$$
-
\int_0^\tau
\gamma ^d \sum _{x,e}
\left(\text{Av}_{\epsilon^{-1}\gamma, x}
(\pi_0, \eta _t)\right)^2 \left[
\beta
(\rho _{\gamma ,t}*e \cdot\nabla J)  \right]
(e \cdot\nabla \phi) (\gamma x,t) \text{d}t
\bigg\vert >\delta\bigg)
\(long)
$$
$$
\le \lim_{\epsilon \rightarrow 0}
\limsup_{\gamma \rightarrow 0}
\Prob_\gamma ^{\mu_\gamma}
\bigg( \beta\Vert\nabla \phi \Vert_{C^0}  \Vert J \Vert_{C^1}
$$
$$
\times \int_0^\tau \gamma ^d \sum _{x,e}\Big\vert
\text{Av}_{\epsilon^{-1}\gamma, x}
(\pi_0 \pi_e, \eta _t)-
\left(\text{Av}_{\epsilon^{-1}\gamma, x}(\pi_0, \eta _t) \right)^2
\Big\vert \text{d}t >\delta \bigg)
=0
$$
Take the limit in $\gamma$ along a convergent subsequence to obtain from
\(mart1), \(oralapl),
\(g2anddoob)  and \(long) that
for every $\delta >0$
with
$\Prob _{\gamma}^{\mu_\gamma}-$probability
going to 1 as $\epsilon $ goes to zero,
 the limit point $\rho$ verifies the inequality
$$
\align
\Big\vert
\int _{T^d}
\rho (r, \tau)\phi( r, \tau)\text{d}r
-\int _{T^d}
\rho (r, 0)\phi_0( r)\text{d}r
- \int\int_{Q_\tau}
\rho(r,t) (\partial_t \phi +\Delta \phi )
\text{d}r \text{d}t
\\
- \int \int_{Q_\tau}
\nabla \phi(r,t)
(\rho -(\alpha _\epsilon *\rho)^2)
\left[ \beta
(\rho*\nabla J) \right]\text{d}r \text{d}t
\Big\vert <\delta
\endalign
$$
in which $\alpha _\epsilon = (2\epsilon)^{-d} \chi_{[-\epsilon,
+\epsilon]^d}$. Let $\epsilon$ go to zero and by the
arbitrariness
of $\delta$ we obtain that $\rho$ is a weak solution of
\(maineq).
To complete the proof of Proposition 1 we have to show that
there is only one weak solution.

Let us then consider $\rho_1$ and $\rho_2$ weak solutions of
\(maineq) with the same initial datum.
Set $w=\rho _1 -\rho_2$. For each $(r_0, t_0) \in Q_\tau$
and each $\epsilon>0$
let us define
$$
\phi_{r_0, t_0, \epsilon }
(r,t)= \theta (t_0 +\epsilon -t , r-r_0)
\(test)
$$
in which $\theta(r,t)= \sum_{x \in \Z ^d} {\Cal G}_t (r-x)$
is the periodic heat kernel, which solves $\partial _t \theta=
\Delta \theta$ on ${T^d}\times (0, \infty)$.
So the test function $\phi_{r_0, t_0, \epsilon }
\in C^\infty (Q_{t_0})$ solves
$$
\partial_t \phi_{r_0, t_0, \epsilon } (r,t)
+\Delta \phi_{r_0, t_0, \epsilon } (r,t)=0
\(invheat)
$$
for all $(r,t) \in Q_{t_0}$.
By using  \(invheat)
we obtain ($\sigma (\rho) =\beta \rho(1-\rho)$)
$$
\int_{T^d} \phi_{r_0, t_0, \epsilon } (r,t_0) w(r,t_0)
\text{d} r=
$$
$$
 \int \int_{Q_{t_0}}
\nabla \phi _{r_0, t_0, \epsilon} (r,t)
\big[(\nabla J*w)(r,t)
\sigma (\rho_1(r,t)) +
(\nabla J*\rho _2)(r,t)\sigma ^\prime(\tilde{\rho}) w(r,t)
\big]
\text{d} r \text{d} t
$$
in which $\tilde {\rho} \in [(\rho_1 \wedge \rho_2)\vee 0 ,
(\rho_1 \vee \rho_2)\wedge 1]$. Since $\sigma(\rho_1) \in [0,\beta/4]$ and
$ \vert \sigma ^\prime(\tilde{\rho})\vert \le \beta$ we obtain
that
$$
\left\vert
\int_{T^d} \phi _{r_0, t_0, \epsilon} (r,t_0)
w(r, t_0) \text{d} r \right\vert
$$
$$
\le{c } [\beta
\Vert \nabla J \Vert _{1} ]
 \esssup_{(r,t) \in Q_{t_0}}
\vert w(r,t)\vert
\int \int _{Q_{t_0}}
\vert \nabla \phi _{r_0, t_0, \epsilon} (r,t)\vert
\text{d} r \text{d} t
\(mnbv)
$$
The integral term on the right hand side
of \(mnbv) is bounded by $c (\sqrt{t_0+\epsilon }
-\sqrt{\epsilon})$,
where $c$ is a constant depending only on the dimension $d$.
By observing that $\phi _{r_0, t_0, \epsilon}(\cdot, t_0)$ is an
approximate identity in $\epsilon$,
we obtain  that there exists a constant $C$ (which depends
only on the $L^1$ norm of $\nabla J$ and on $d$) such that
for almost every $(r_0 ,t_0) \in Q_{\tau}$ we have
that
$$
\vert w(r_0, t_0) \vert
\le C \sqrt{t_0}
\esssup_{(r,t) \in Q_{t_0}}
\vert w(r,t)\vert
$$
and so
$$
\esssup_{(r,t) \in Q_{\tau}}\vert w(r, t) \vert
\le C \sqrt{\tau}
\esssup_{(r,t) \in Q_{\tau}}
\vert w(r,t)\vert.
$$
Choosing $\tau$ such that $C\sqrt{\tau} <1$ implies
local uniqueness in time. But $C$ depends only on
$J$ and $d$, so
the global uniqueness follows by a bootstrap argument.
\qed
\vskip 0.3 cm

\proclaim{Proposition 1}
If $\rho_0\in C^2(T^d)$, then the solution
$\rho$ of \(weakform) can be chosen in
$C^{2,1} (T^d \times {\Bbb R}^+)$.
\endproclaim

\vskip 0.3 cm

{\it Proof:}
given a solution of \(weakform) define
$$
F (r,t)= (\nabla J * \rho_t) (r)
\(defF)
$$
so $F \in C^{\infty , 0} (T^d \times {\Bbb R}^+; {\Bbb R}^d)$.
Observe that this means that
$\rho$ solves in the weak sense analogous to  \(weakform)
the equation
$$
\partial _t \rho =
\nabla \cdot \left\{
\nabla \rho - \beta \rho (1-\rho) F \right\}
\(parabolic)
$$
with initial condition $\rho_0 \in C^2$.
But \(parabolic) is a nondegenerate parabolic equation
which has a classical solution (Chapter 7, Section 4 of [\rcite{Friedman}]),
besides having a unique weak solution  by the same
argument used to prove uniqueness in the proof of Theorem 2.
\qed

\head
Acknowledgements
\endhead
We would like to thank A. Asselah, O. Benois,
L. Bonaventura, J. Percus, J. Quastel,
 H. Spohn, S.R.S. Varadhan,
L. Xu, H.T. Yau and in particular  E. Presutti
for very fruitful discussions.
This work was partially supported by NSF--DMR
92--13424 4--20946 and AFOSR 0159 4--26435 (J.L.L.)
and by the Swiss National Science Foundation, Project  20--$41^\prime925.94$
 (G.G.).

\enddocument